\documentclass{article}
\usepackage{spconf,amsmath,graphicx}
\usepackage{amssymb}
\usepackage{amscd}

\def\expk{{e^{-i \frac{2\pi}{T}k t}}}

\title{FOURIER DOMAIN BEAMFORMING FOR MEDICAL ULTRASOUND}
\name{Tanya Chernyakova, Yonina C. Eldar, Ron Amit}
\address{Technion - Israel Institute of Technology\\
Dept. of Electrical Engineering\\
Haifa 32000, Israel}

\begin{document}
\ninept
\maketitle
\begin{abstract}
Sonography techniques use multiple transducer elements for tissue visualization. Signals detected at each element are sampled prior to digital beamforming. The required sampling rates are up to 4 times the Nyquist rate of the signal and result in considerable amount of data, that needs to be stored and processed. A developed technique, based on the finite rate of innovation model, compressed sensing (CS) and Xampling ideas, allows to reduce the number of samples needed to reconstruct an image comprised of strong reflectors. A significant drawback of this method is its inability to treat speckle, which is of significant importance in medical imaging.
Here we build on previous work and show explicitly how to perform beamforming in the Fourier domain.
Beamforming in frequency exploits the low bandwidth of the beamformed signal and allows to bypass the oversampling dictated by digital implementation of beamforming in time.
We show that this allows to obtain the same beamformed image as in standard beamforming but from far fewer samples. Finally, we present an analysis based CS-technique that allows
for further reduction in sampling rate, using only a portion of the beamformed signal's bandwidth, namely, sampling the signal at sub-Nyquist rates. We demonstrate our methods on in vivo cardiac ultrasound data and show that reductions up to 1/25 over standard beamforming rates are possible.%and up to 1/3 of Nyquist rate are possible.
%In our work we perform the beamforming process in frequency domain and exploit the bandwidth of beamformed signal to reconstruct the image perfectly from a much smaller number of samples. We also incorporate a novel CS technique, referred to as analysis, for further reduction in the number of samples, while using the partial bandwidth of the beamformed signal. We demnstrate the success of our methods on in vivo cardiac ultrasound data.
\end{abstract}
\section{Introduction}
\label{sec:intro}
Diagnostic ultrasound has been used for decades to visualize body structures. The overall imaging process is described as follows: An energy pulse is transmitted along a narrow beam. During its propagation echoes are scattered by acoustic impedance perturbations in the tissue, and detected by the elements of the transducer. Collected data are sampled and digitally processed in a way referred to as beamforming, which results in signal-to-noise ratio (SNR) enhancement. Such a beamformed signal forms a line in the image.

According to the classic Shannon-Nyquist theorem \cite{shannon1949communication}, the sampling rate at each transducer element should be at least twice the bandwidth of the detected signal. Rates up to 4 times the Nyquist rate are required in order to avoid artifacts caused by digital implementation of beamforming \cite{szabo2004diagnostic}.
Taking into account the number of transducer elements and the number of lines in an image, the amount of sampled data that needs to be digitally processed is enormous, motivating methods to reduce sampling rates. %or, alternatively, the compression of sampled data.
Reduction of processing rate is possible within the classical sampling framework by exploiting the fact that the signal is modulated onto a carrier and occupies only a portion of its entire baseband bandwidth. Accordingly, modern systems digitally down-sample the data at the system's front-end. However, this does not change the sampling rate since demodulation takes place in the digital domain. In addition, the processing rate may be reduced up to 1/4 of the sampling rate, but the signal becomes complex in this setup, and the number of samples effectively is only twice smaller.

A different approach to sampling rate reduction is introduced in \cite{tur2011innovation}. Tur et. al. regard the ultrasound signal detected by each receiver within the framework of finite rate of innovation (FRI) \cite{vetterli2002sampling}, modeling it as $L$ replicas of a known pulse-shape, caused by scattering of the transmitted pulse from reflectors, located along the transmitted beam. Such an FRI signal is fully described by $2L$ parameters, corresponding to the replica's delays and amplitudes. These parameters can be extracted from a small set of the signal's Fourier series coefficients. A mechanism, referred to as Xampling, derived in \cite{gedalyahu2011multichannel}, extracts such a set of coefficients from $4L$ real-valued samples. This work is continued in \cite{wagner2012compressed}, where Wagner et. al. introduce a generalized scheme, referred to as compressed beamforming, which allows to compute the Fourier series coefficients of the beamformed signal from the low-rate samples of signals detected at each element. The problem of reconstruction of the beamformed signal from a small number of its Fourier series coefficients is solved via a compressed sensing (CS) technique, while assuming a small number $L$ of replicas. This approach allows to reconstruct an image comprised of macroscopic perturbations, but cannot treat the speckle, which is of significant importance in medical imaging.

In our work we extend the notion of compressed beamforming to beamforming in frequency and show explicitly how to perform it. Beamforming in frequency exploits the low bandwidth of the beamformed signal and allows to bypass the oversampling dictated by the digital implementation of beamforming in time. We reconstruct the beamformed signal perfectly with a simple inverse discrete Fourier transform (IDFT) from a small set of its discrete Fourier transform (DFT) coefficients, that are computed from low-rate samples of individual signals. We show that beamforming in frequency allows to preserve the integrity of an image with 7 fold reduction in the number of samples used for its reconstruction.

Finally, we introduce an analysis based CS-technique \cite{candes2011compressed} for signal reconstruction using only a portion of the beamformed signal's bandwidth. We demonstrate our methods on in vivo cardiac ultrasound data and show that reductions of up to 1/25 of beamforming rate and up to 1/3 of Nyquist rate are possible.
The proposed reconstruction method outperforms the classic synthesis CS approach used in \cite{wagner2012compressed}, when the same number of samples is used.

The rest of the paper is organized as follows: in Section \ref{sec:beam in time}, we review beamforming in time. In Section \ref{sec:beam in freq} we describe the principles of frequency domain beamforming. In Section \ref{sec:analysis} we describe how the reduction in sampling rate is achieved. In Section \ref{sec:CS} we discuss two possible CS approaches to signal reconstruction.
\section{Beamforming in time}
\label{sec:beam in time}
Most modern imaging systems use multiple transducer elements to transmit and receive acoustic pulses. Appropriate processing of the signals detected by the individual array elements allows to enhance the quality of the resulting image. The most commonly used technique, referred to as beamforming, increases SNR by averaging the individual signals after their alignment with appropriate time-delays.
We begin by describing the beamforming process which takes place in a typical B-mode imaging cycle. Our presentation is based mainly on \cite{jensen1999linear} and \cite{wagner2012compressed}.
%\[h]
\begin{figure}[htb]
\begin{minipage}[b]{1.0\linewidth}
  \centering
 \vspace{-0.75cm}
  \centerline{\includegraphics[width=4.5cm]{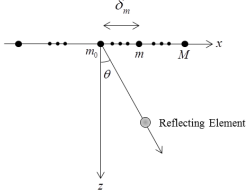}}
  \vspace{-0.3cm}
\end{minipage}
\caption{$M$ receivers aligned along the $x$ axis. An acoustic pulse is transmitted at direction $\theta$. The echoes scattered from perturbation in radiated tissue are received by the array elements.}
\label{fig:array}
\end{figure}

Consider an array comprised of $M$ transceiver elements aligned along the $x$ axis, as illustrated in Fig. \ref{fig:array}. The reference element $m_0$ is set at the origin and the distance to the $m$-th element is denoted by $\delta_m$. The image cycle begins at $t=0$, when the array transmits an energy pulse in the direction $\theta$. The pulse propagates trough the tissue at speed $c$, and at time $t\geq0$ its coordinates are $(x,z)=(ct\sin{\theta},ct\cos{\theta})$. A potential point reflector located at this position scatters the energy, such that the echo is detected by all array elements at a time depending on their locations. Denote by $\varphi_m(t;\theta)$ the signal detected by the $m$-th element and by $\hat{\tau}_m(t;\theta)$ the time of detection. It is readily seen that:
\begin{equation}
\label{tau^m}
\hat{\tau}_m(t;\theta)=t+\frac{d_m(t;\theta)}{c},
\end{equation}
where $d_m(t;\theta)=\sqrt{(ct\cos{\theta})^2+(\delta_m-ct\sin{\theta)}^2}$ is the distance traveled by the reflection.
Beamforming involves averaging the signals detected by multiple receivers while compensating the differences in detection time.

Using \eqref{tau^m}, the detection time at $m_0$ is $\hat{\tau}_{m_0}(t;\theta)=2t$ since $\delta_{m_0}=0$. Applying an appropriate delay to $\varphi_m(t;\theta)$, such that the resulting signal $\hat{\varphi}_m(t;\theta)$ satisfies $\hat{\varphi}_m(2t;\theta)=\varphi_m(\hat{\tau}_m(t;\theta))$, we can align the reflection detected by the $m$-th receiver with the one detected at $m_0$. Denoting $\tau_m(t;\theta)=\hat{\tau}_m(t/2;\theta)$  and using \eqref{tau^m}, the following aligned signal is obtained:
\begin{align}
\label{phim}
\hat{\varphi}_m(t;\theta)&=\varphi_m(\tau_m(t;\theta);\theta),\\ \nonumber
\tau_m(t;\theta)&=\frac{1}{2}\left(t+\sqrt{t^2-4(\delta_m/c)t\sin{\theta}+4(\delta_m/c)^2}\right).
\end{align}
The beamformed signal may now be derived by averaging the aligned signals:
\begin{equation}
\label{phi beamformed}
\Phi(t;\theta)=\frac{1}{M}\sum_{m=1}^M{\hat{\varphi}_m(t;\theta)}.
\end{equation}

Ultrasound systems perform the beamforming process defined in \eqref{phi beamformed} in the digital domain, implying that the analog signals $\varphi_m(t;\theta)$ detected at the receiver elements are first sampled. Rates up to 4 times the Nyquist rate, dictated by the bandwidth of the individual signal, are required in order to improve the system's beamforming resolution and to avoid artifacts caused by digital implementation. From now on we will denote this sampling rate as the beamforming sampling rate $f_s$.

To conclude this section we evaluate the number of samples taken at each transducer element. Our evaluation is based on the imaging setup used to acquire in vivo cardiac data. The acquisition was performed with a GE breadboard ultrasonic scanner of 64 acquisition channels. The radiated depth $r=16$ cm and the speed of the sound $c=1540$ m/sec yield a signal of duration $T=2r/c\simeq210$ $\mu$sec. The acquired signal is characterized by a narrow bandpass bandwidth of $2$ MHz, centered at the carrier frequency $f_0\approx3.1$ MHz, leading to a beamforming rate of $f_s\approx16$ MHz and $Tf_s=3360$ real-valued samples.
%To conclude this section we evaluate the number of samples that needs to be taken to form a typical cardiac image. Our evaluation is based on the imaging setup, described in Section \ref{sec:beam in freq}, used to acquire in vivo cardiac data with GE Healthcare imaging system.
%The nominal radiated depth $r=16$ cm and the speed of the sound $c=1540$ m/sec yield a signal of duration $T=2r/c\simeq210$ $\mu$sec. The carrier frequency is $f_0\approx3.1$ Mhz, leading to a beamforming rate of $f_s\approx16$ MHz and $Tf_s=3360$ real-valued samples. For an array of $M=64$ elements and $K=120$ transmitted beams, the overall number of samples is $3360\cdot64\cdot120\simeq25.8\cdot10^6$. Our goal is to reduce this number without compromising image quality.
%
\section{Beamforming in frequency}
\label{sec:beam in freq}
We now show that beamforming can be performed equivalently in the frequency domain, leading to substantial reduction in the number of samples, needed to obtain the same image quality.

We extend the notion of compressed beamforming, introduced in \cite{wagner2012compressed}, to beamforming in frequency and show that a linear combination of the DFT coefficients of the individual signals, sampled at the beamforming rate $f_s$,  yields the DFT coefficients of the beamformed signal, sampled at the same rate. We follow the steps in \cite{wagner2012compressed} and start from the computation of the Fourier series coefficients of the beamformed signal $\Phi(t;\theta)$.

The support of $\Phi(t;\theta)$ is limited to $[0,T]$, with $T$ defined by the transmitted pulse penetration depth. Its Fourier series coefficients are given by:
\begin{equation}
\label{fourier coeff of beamformed 1}
c_k^s=\frac{1}{T}\int_0^T \Phi(t;\theta)\expk dt.
\end{equation}
Plugging \eqref{phi beamformed} into \eqref{fourier coeff of beamformed 1}, it can be shown that
\begin{equation}
\label{fourier coeff of beamformed 2}
c_k^s=\frac{1}{M}\sum_{m=1}^M c_{k,m}^s,
\end{equation}
where $c_{k,m}^s$ are the Fourier coefficients of $\hat{\varphi}_m(t;\theta)$. These coefficients can be written as
\begin{equation}
\label{c k m}
c_{k,m}^s=\frac{1}{T}\int_0^T g_{k,m}(t;\theta) \varphi_m(t;\theta)dt,
\vspace{-0.3cm}
\end{equation}
with
\begin{align}\label{g j m q j m}
g_{k,m}(t;\theta)=&q_{k,m}(t;\theta)\expk, \nonumber \\
q_{k,m}(t;\theta)=& I_{[|\gamma_m|,\tau_m(T;\theta))}(t) \left(1+\frac{\gamma_m^2\cos{\theta}^2}{(t-\gamma_m\sin{\theta})^2}\right)\times  \\*
&\exp{\left\{i\frac{2\pi}{T}k\frac{\gamma_m-t\sin{\theta}}{t-\gamma_m\sin{\theta}}\gamma_m\right\}}, \nonumber
\end{align}
where $\gamma_m=\delta_m/c$, and $I_{[a,b]}$ is the indicator function.

Our next step is to substitute $\varphi_m(t)$ by its Fourier series coefficients. Denoting the $n$-th Fourier coefficient by $\varphi_m^s[n]$ and using \eqref{g j m q j m} we can rewrite \eqref{c k m} as follows:
\begin{equation}\label{c k m fourier}
 c_{k,m}^s=\sum_n \varphi_m^s[n]Q_{k,m;\theta}[k-n],
\end{equation}
where $Q_{k,m;\theta}[n]$ are the Fourier coefficients of $q_{k,m}(t;\theta)$ with respect to $[0,T)$. According to Proposition 1 in \cite{wagner2012compressed}, $c_{k,m}^s$ can be approximated sufficiently well when we replace the infinite summation in \eqref{c k m fourier} by the finite one:
\begin{equation}\label{c k m approx}
c_{k,m}^s\simeq\sum_{n\in\nu(k)}\varphi_m^s[n]Q_{k,m;\theta}[k-n].
\end{equation}
The set $\nu(k)$ is defined according to the decay properties of $\{Q_{k,m;\theta}[n]\}$.
Equations \eqref{fourier coeff of beamformed 2} and \eqref{c k m approx} provide a relationship between the Fourier series coefficients of the beamformed and the individual signals. Denote by $N=\lfloor T\cdot f_s\rfloor$ the number of samples in each signal. Since all signals are sampled at a rate which is higher than their Nyquist rate, the relation between the DFT of length $N$ and the Fourier series coefficients is given by:
\begin{align}\label{DFT-Fourier series}
    c_k=Nc_k^s,~~~ \varphi_m[n]=N\varphi_m^s[n],
\end{align}
where $c_k$ and $\varphi_m[n]$ denote the DFT coefficients of the beamformed and individual signals respectively. Plugging \eqref{DFT-Fourier series} into \eqref{fourier coeff of beamformed 2}  and \eqref{c k m fourier}, we get the desired relation:
\begin{align}\label{DFT beam - DFT individual}
    c_k&\simeq\frac{1}{M}\sum_{m=1}^M\sum_{n\in\nu(k)}\varphi_m[n]Q_{k,m;\theta}[k-n].
\end{align}
Note that in order to calculate an arbitrary set $\kappa$ of DFT coefficients of the beamformed signal, we need $\nu=\cup_{k\in\kappa}\nu(k)$ DFT coefficients of each one of the individual signals.

\begin{figure}[htb]
\begin{minipage}[b]{0.5\linewidth}
  \centering
  \centerline{\includegraphics[width=4.6cm]{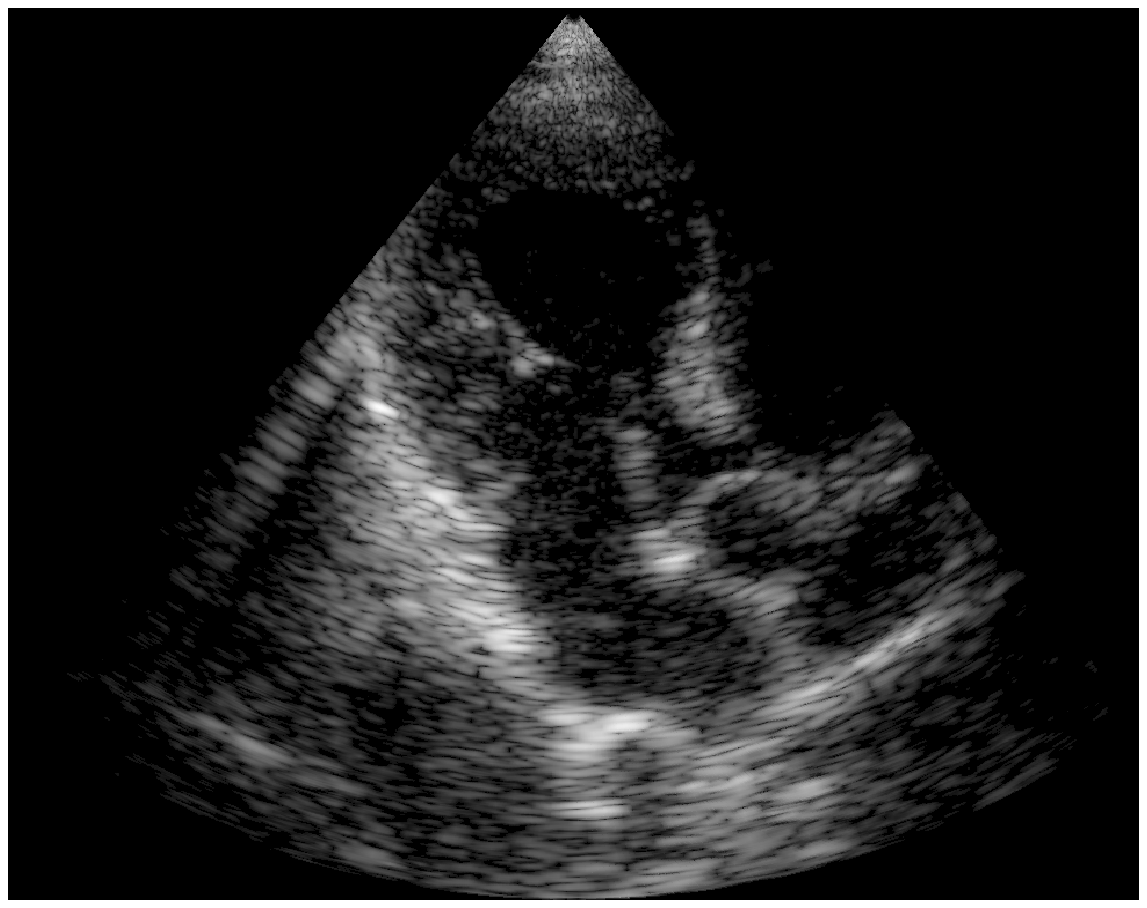}}
\vspace{-0.2cm}
  \centerline{(a)}\medskip
  \vspace{-0.5cm}

\end{minipage}
\hfill
\begin{minipage}[b]{0.5\linewidth}
  \centering
  \centerline{\includegraphics[width=4.6cm]{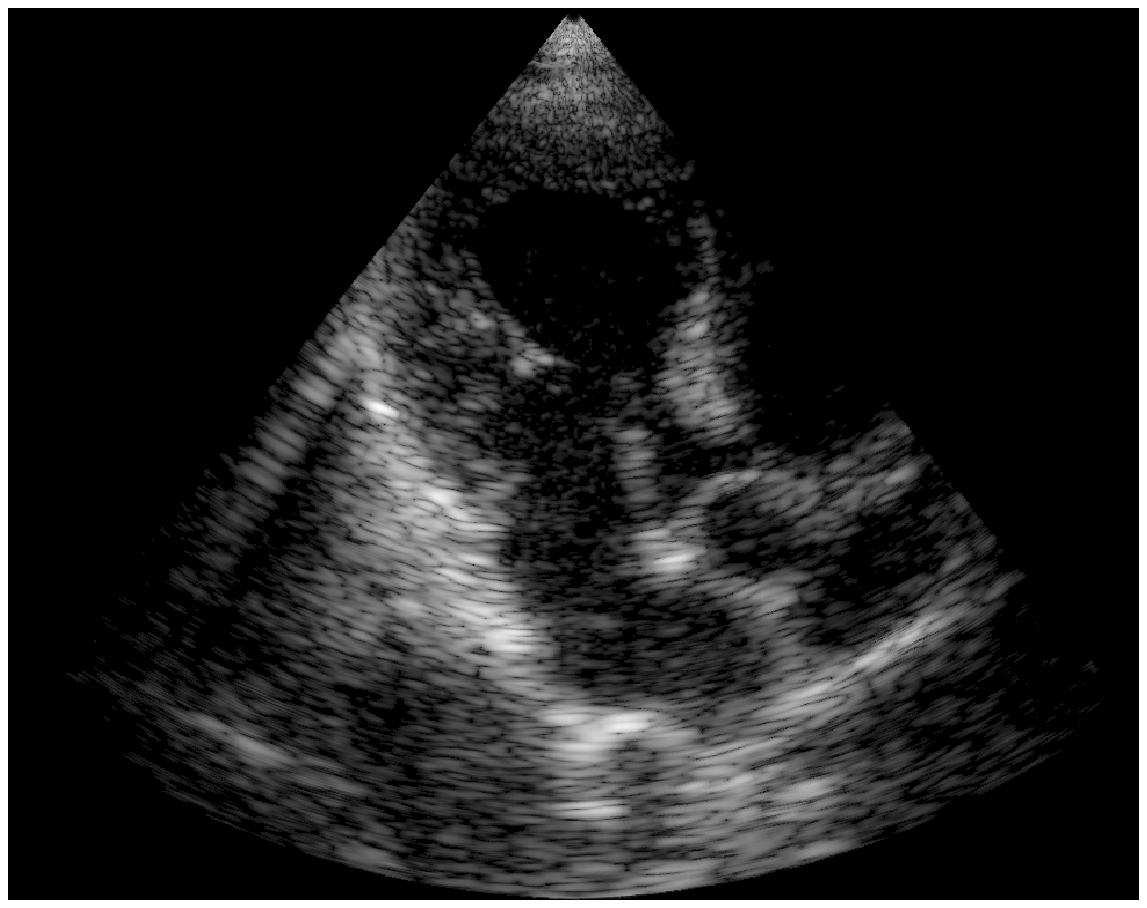}}
\vspace{-0.2cm}
  \centerline{(b)}\medskip
  \vspace{-0.5cm}

\end{minipage}
\caption{Cardiac images constructed with different beamforming techniques. (a) Time domain beamforming. (b) Frequency domain beamforming. }
\label{fig:time vs freq beamformed}
\end{figure}

Applying an IDFT on $\{c_k\}_{k=1}^N$, we obtain the beamformed signal. We can now proceed to standard image generation steps which include log-compression and interpolation.
To demonstrate the equivalence of beamforming in time and frequency, we applied both methods on in vivo cardiac data obtained as explained in Section \ref{sec:beam in time}, yielding the images shown in Fig. \ref{fig:time vs freq beamformed}. As can be seen, both images are identical.
\vspace{-0.3cm}
\vspace{-0.5cm}

\section{Rate Reduction by Beamforming in Frequency}
\label{sec:analysis}
In the previous section we showed the equivalence of beamforming in time and frequency. We next demonstrate that beamforming in frequency allows to reduce the required number of samples of the individual signals. To this end we consider two questions: 1) how many DFT coefficients of the beamformed signal do we need for its perfect reconstruction; 2) how many samples of the individual signals should be taken in order to compute these DFT coefficients?
\subsection{Parametric representation}
\label{ssec:param representation}
We begin by answering the first question using a parametric model for the beamformed signal.
According to \cite{tur2011innovation, wagner2012compressed}, the beamformed signal can be modeled as a sum of a small number of replicas of the known transmitter pulse with unknowns amplitudes and delays:
\begin{equation}\label{beam FRI}
    \Phi(t;\theta)\simeq\sum_{l=1}^L\tilde{b}_l h(t-t_l),
\end{equation}
where $h(t)$ is the transmitted pulse-shape, $L$ is the number of scattering elements in direction $\theta$, $\{\tilde{b}_l\}_{l=1}^L$ are the unknown amplitudes of the reflections and $\{t_l\}_{l=1}^L$ denote the times at which the reflection from the $l$-th element arrived at the reference receiver $m_0$. Sampling both sides of \eqref{beam FRI} at rate $f_s$ and quantizing the unknown delays $\{t_l\}_{l=1}^L$ with quantization step $1/f_s$, such that $t_l=q_l/f_s, q_l\in\mathbb{Z}$, we can rewrite \eqref{beam FRI} as follows:
\begin{equation}\label{beam FRI n}
    \Phi[n;\theta]\simeq\sum_{l=1}^L\tilde{b}_l h[n-q_l]=\sum_{l=0}^{N-1} b_l h[n-l],
\end{equation}
where
\vspace{-0.5cm}
\begin{align}
\label{tilde b - b}
b_l = \left\{ \begin{array}{rl}
 \tilde{b}_l &\mbox{ if $l=q_l$} \\
  0 &\mbox{ otherwise}.
       \end{array} \right.
\end{align}
Calculating the DFT using \eqref{beam FRI n}:
\begin{align}
\label{DFT BF}
c_k=\sum_{n=0}^{N-1}\Phi[n;\theta]e^{-i \frac{2\pi}{N}k n}=h_k\sum_{l=0}^{N-1} b_l e^{-i \frac{2\pi}{N}k l},
\end{align}
where $h_k$ is the DFT coefficient of $h[n]$.

\begin{figure}[htb]
\begin{minipage}[b]{1.0\linewidth}
  \centering
  \centerline{\includegraphics[width=5.0cm]{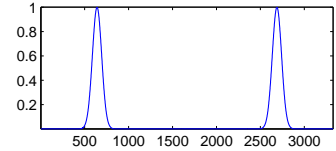}}
\end{minipage}
\vspace{-0.6cm}
\caption{The DFT coefficients of the transmitted pulse. The bandwidth of the waveform is $2$ MHz, $f_0=3.1$ MHz, $f_s=16$ Mhz.}
\label{fig:pulseDFT}
\vspace{-0.5cm}
\end{figure}

The transmitted pulse $h(t)$ may be modeled as a narrowband waveform, $g(t)$, modulated by a carrier at frequency $f_0$: $h(t)=g(t)\cos(2\pi f_0 t)$. When such a pulse is sampled at rate $f_s$, most of its DFT coefficients are zero, as shown in Fig. \ref{fig:pulseDFT}. Obviously, \eqref{DFT BF} implies that the only non-zero DFT coefficients are in the bandwidth of the transmitted pulse. This allows us to exploit the low bandwidth of the beamformed signal and calculate only non-zero DFT coefficients.
Since $\Phi[n;\theta]$ is real, its DFT coefficients are symmetric, thus we only need to know half of the overall non-zero elements. Denote the set of non-zero DFT coefficients by $\kappa$. In typical cardiac imaging the bandwidth of $g(t)$ is equal to $2$ MHz, the modulation frequency $f_0=3.1$ MHz, and the sampling rate $f_s=16$ MHz, leading to $K=|\kappa|\approx360$. Once these $K$ coefficients are known, we can reconstruct $\Phi[n;\theta]$ by padding the elements of $\kappa$ with an appropriate number of zeros and performing an IDFT. Hence, we have shown that the number of DFT coefficients of the beamformed signal required for its perfect reconstruction is the cardinality of the set $\kappa$. This number is only $1/9$ of the overall number of DFT coefficients of beamformed signal dictated by $f_s$.
\vspace{-0.1cm}
\subsection{Reduced rate sampling}
\label{ssec:red rate samp}
We now address the second question: how many samples of the individual signals should be taken in order to compute the set of non-zero DFT coefficients $\kappa$?

\begin{figure}[htb]
\begin{minipage}[b]{0.5\linewidth}
  \centering
  \centerline{\includegraphics[width=4.6cm]{plots/timeBeamformed}}
\vspace{-0.2cm}
  \centerline{(a)}\medskip
\vspace{-0.5cm}
\end{minipage}
\hfill
\begin{minipage}[b]{0.5\linewidth}
  \centering
  \centerline{\includegraphics[width=4.6cm]{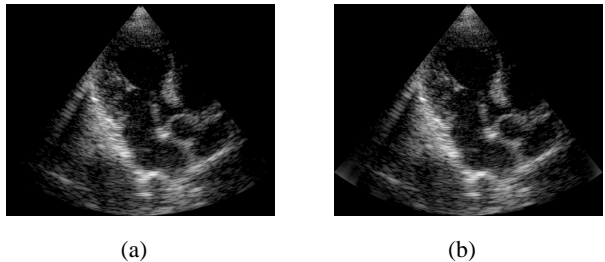}}
\vspace{-0.2cm}
  \centerline{(b)}\medskip
     \vspace{-0.5cm}

\end{minipage}
\caption{Cardiac images constructed with different beamforming techniques. (a) Time domain beamforming, $3360$ samples per individual signal. (b) Frequency domain beamforming using $480$ samples per individual signal (7 fold reduction in sampling rate).}
\label{fig:time vs freq 360 beamformed}
\end{figure}

As shown in Section \ref{sec:beam in freq}, we need to know a set $\nu$ of the DFT coefficients of each individual signal in order to compute a set $\kappa$ of the DFT coefficients of the beamformed signal. The experimental results show that in ultrasound imaging scenario, $|\kappa|\leq|\nu|\leq1.33|\kappa|$.
According to \eqref{DFT-Fourier series}, we equivalently seek the same set $\nu$ of Fourier series coefficients of each individual signal. The above problem is addressed in \cite{wagner2012compressed}. A mechanism, proposed there, allows to obtain a set $\kappa$ of Fourier coefficients of the beamformed signal from $|\nu|$ samples of each of the individual signals, filtered with an appropriate kernel. Namely, the number of the samples taken from the individual signal is $|\nu|$.
For the setup mentioned in Subsection \ref{ssec:param representation}, $|\kappa|\approx360$, implying that in the worst case we need only $|\nu|=1.33|\kappa|\approx480$ samples of the individual signal, while for beamforming in time domain we need $\approx3360$ samples as shown in Section \ref{sec:beam in time}. This allows us to achieve 7-fold reduction in sampling rate without compromising image quality.
The images, created by these two techniques, shown in Fig. \ref{fig:time vs freq 360 beamformed}, are identical.
%
%Note that the reduction in sampling rate stems from exploiting the low bandwidth of the beamformed signal by taking its $|\kappa|$ non-zero DFT coefficients and from the ability to calculate this set from low-rate samples of the individual signals.
%
\vspace{-0.3cm}
\vspace{-0.7cm}
\section {Further Reduction via CS}
\label{sec:CS}
We showed that it is possible to reconstruct a beamformed signal perfectly from the set $\kappa$ of its non-zero DFT coefficients, computed from a small number of samples of the individual signals. We show next that further reduction in sampling rate is possible, when taking only a subset $\mu\subset\kappa$, $|\mu|=M < K=|\kappa|$, of DFT coefficients of the beamformed signal.
%\subsection{Problem formulation}
%\label{ssec:problem form}

Defining a $K$-length vector $\mathbf{c}$ with $k$-th entry $c_k/h_k, k\in\kappa$, we can rewrite \eqref{DFT BF} in matrix form:
\begin{equation}\label{DFT BF matrix full}
\mathbf{c}=\mathbf{D} \mathbf{b}
\end{equation}
where $\mathbf{D}$ is a $K\times N$ matrix formed by taking the set $\kappa$ of rows from an $N\times N$ DFT matrix, and vector $\mathbf{b}$ is of length $N$ with $l$-th entry $b_l$.
Since from now on only subset $\mu$ is given, define an $M$-length vector $\mathbf{c_{\mu}}$ with $k$-th entry $c_k/h_k, k\in\mu$ and rewrite \eqref{DFT BF matrix full} as follows:
\begin{equation}\label{DFT BF matrix}
\mathbf{c_{\mu}}=\mathbf{A} \mathbf{D} \mathbf{b}
\end{equation}
where $\mathbf{A}$ is $M\times K$ measurement matrix which picks the subset $\mu$ of rows from $\mathbf{D}$, implying that $\mathbf{A D}$ is $M\times K$ matrix formed by taking the set $\mu$ of rows from an $N\times N$ DFT matrix.This property of $\mathbf{A D}$ will be used below.
\subsection{Synthesis approach}
Since from \eqref{beam FRI n} the signal of interest is completely defined by the unknown delays and amplitudes, a possible approach is to extract those values from the available set $\mu$ of DFT coefficients.
Equation \eqref{DFT BF matrix} can be viewed from a CS perspective, when we assume that the number of scatterers $L$ is small, since according to \eqref{tilde b - b} it implies that the vector  $\mathbf{b}$ is sparse. Hence \eqref{DFT BF matrix} has the form of a classic sparse synthesis model \cite{eldar2012compressed}, where vector $\mathbf{c}_{\mu}$ has a sparse representation in $\mathbf{AD}$. The goal is to reconstruct an $L$-sparse vector $\mathbf{b}$ from its projection onto a subset of $M$ orthogonal vectors given by the rows of matrix $\mathbf{AD}$. With an appropriate choice of $L$ and subset of fourier coefficients $\mu$, such a problem can be solved using CS methodology, including $l1$ optimization and greedy algorithms.
Indeed, in \cite{wagner2012compressed}, orthogonal matching pursuit (OMP) \cite{tropp2007signal} shows sufficiently good performance.

The synthesis approach has a significant drawback. The assumption of a small number of reflecting elements $L$, forces us to treat only the strong reflectors located in direction $\theta$. In such a setup we essentially loose all the weak reflectors that appear as speckle, namely, granular pattern that can be seen in Fig. \ref{fig:time vs freq beamformed}, and carry important information in medical imaging.
\subsection{Analysis approach}
To avoid loss of speckle information, imposed by the assumption of sparsity, we propose using an analysis approach \cite{candes2011compressed}. In this methodology we aim to reconstruct the set $\kappa$ from its subset $\mu$, while assuming that the analyzed vector $\mathbf{D}^*\mathbf{c}$ is compressible. The analysis approach can be translated into the $\textit{l}_1$ optimization problem:
\begin{equation}\label{l1 sol}
    \min_{\mathbf{c}}\|\mathbf{D}^* \mathbf{c}\|_1  \textrm{~~~subject to~~~} \|\mathbf{A}\mathbf{c}-\mathbf{c}_\mu\|_2\leq\varepsilon.
\end{equation}
%\begin{equation}\label{l1 sol}
%    \min_{\mathbf{c}_\kappa}\|\mathbf{D}^* \mathbf{c}_\kappa\|_1  \textrm{  subject to  } \|\mathbf{A}\mathbf{c}_\kappa-\mathbf{c}\|_2\leq\varepsilon
%\end{equation}
%
According to Theorem 1.4 in \cite{candes2011compressed}, the solution to \eqref{l1 sol} is very accurate, if the measurement matrix $\mathbf{A}$ satisfies the restricted isometry property adapted to $\mathbf{D}$ (D-RIP) and the elements of $\mathbf{D}^* \mathbf{c}$ decay rapidly. As we mentioned before, $\mathbf{AD}$ is a partial DFT matrix, therefore, according to results in \cite{candes2011compressed}, $\mathbf{A}$ satisfies the D-RIP.

A typical beamformed ultrasound signal is comprised of a relatively small number of strong reflections and a bunch of much weaker scattered echoes. It is, therefore, natural to assume that $\mathbf{b}$ is compressible, implying that $\mathbf{c}$ has a compressible expansion in $\mathbf{D}$. Since $\mathbf{D}$ is a partial DFT matrix, its Gram matrix is nearly diagonal, implying that $\mathbf{D}^* \mathbf{c}$ is also compressible \cite{candes2011compressed} and satisfies the decay requirement.

\begin{figure}[htb]
\begin{minipage}[b]{0.5\linewidth}
  \centering
  \centerline{\includegraphics[width=4.6cm]{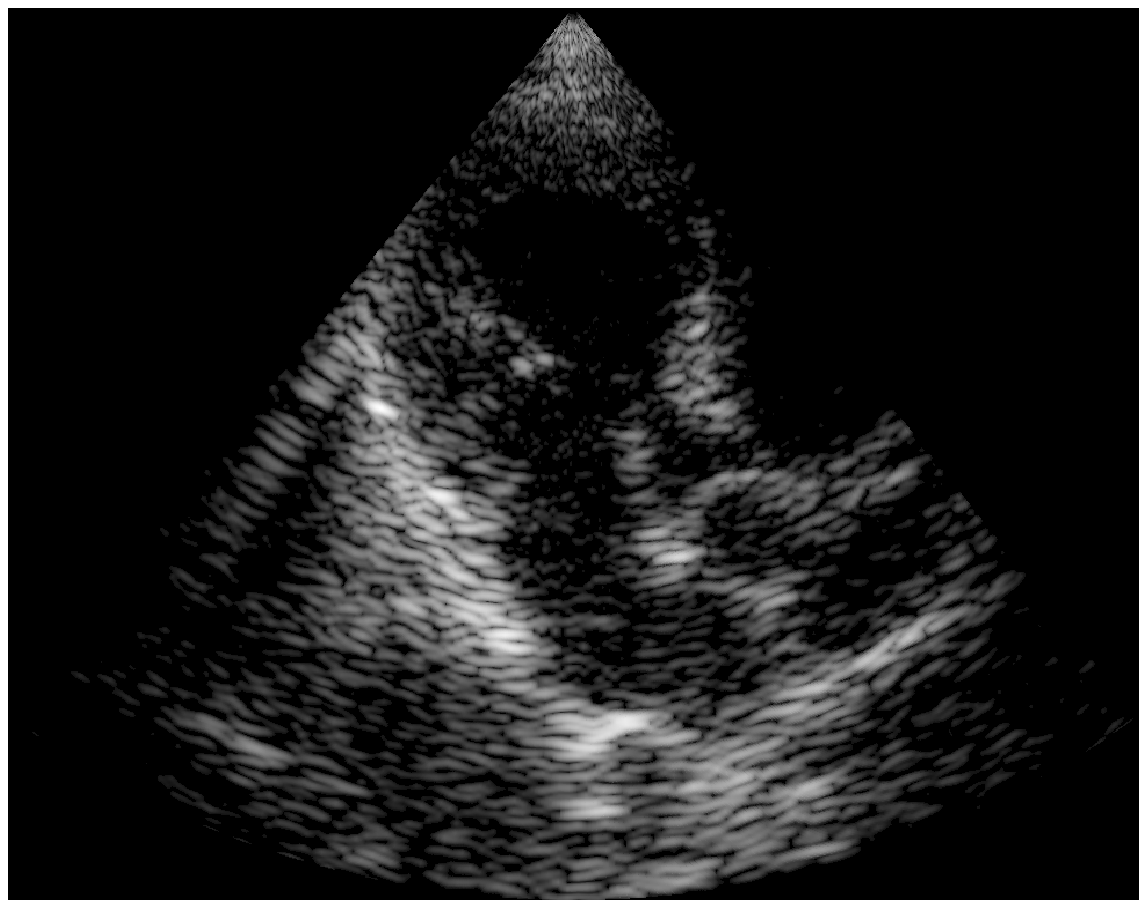}}
\vspace{-0.2cm}
  \centerline{(a)}\medskip
    \vspace{-0.5cm}
\end{minipage}
\hfill
\begin{minipage}[b]{0.5\linewidth}
  \centering
  \centerline{\includegraphics[width=4.6cm]{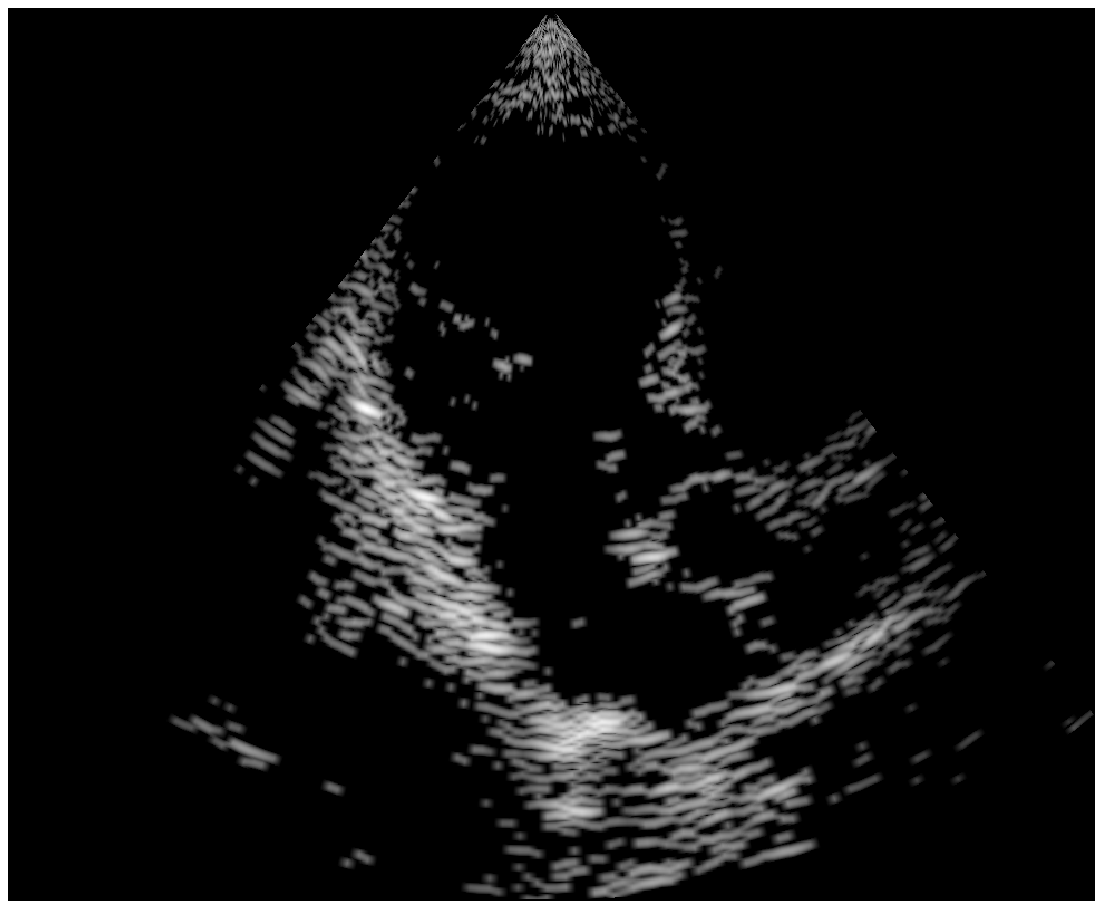}}
\vspace{-0.2cm}
  \centerline{(b)}\medskip
    \vspace{-0.5cm}
\end{minipage}
\caption{Cardiac images constructed from partial spectrum data with 25 fold reduction in sampling rate. (a) Modified $\textit{l}_1$ optimization solution. (b) OMP based reconstruction. }
\label{fig:part BW}
\end{figure}

To demonstrate the proposed method, a subset $\mu$ of $100$ Fourier coefficients corresponding to the central frequency samples in the bandwidth of the transmitted pulse were chosen. To calculate $\mu$ we need at most $133$ samples per individual signal, implying 25 fold reduction in sampling rate. The result is shown in Fig. \ref{fig:part BW} (a). To compare the proposed solution with the previously developed OMP based method \cite{wagner2012compressed}, the same subset $\mu$ was used to reconstruct the beamformed signal assuming $L=25$ strong reflectors in each direction $\theta$. The resulting image is shown in Fig. \ref{fig:part BW} (b).
\vspace{-0.3cm}
\vspace{-0.7cm}

%% Below is an example of how to insert images. Delete the ``\vspace'' line,
%% uncomment the preceding line ``\centerline...'' and replace ``imageX.ps''
%% with a suitable PostScript file name.
%% -------------------------------------------------------------------------
%\begin{figure}[htb]
%
%\begin{minipage}[b]{1.0\linewidth}
%  \centering
%  \centerline{\includegraphics[width=8.5cm]{plots/image1}}
%%  \vspace{2.0cm}
%  \centerline{(a) Result 1}\medskip
%\end{minipage}
%%
%\begin{minipage}[b]{.48\linewidth}
%  \centering
%  \centerline{\includegraphics[width=4.0cm]{plots/image3}}
%%  \vspace{1.5cm}
%  \centerline{(b) Results 3}\medskip
%\end{minipage}
%\hfill
%\begin{minipage}[b]{0.48\linewidth}
%  \centering
%  \centerline{\includegraphics[width=4.0cm]{plots/image4}}
%%  \vspace{1.5cm}
%  \centerline{(c) Result 4}\medskip
%\end{minipage}
%%
%\caption{Example of placing a figure with experimental results.}
%\label{fig:res}
%%
%\end{figure}

%\vfill
%\pagebreak

\vfill\pagebreak

\bibliographystyle{IEEEbib}
\bibliography{general}

\end{document}